\documentclass[aps,prx,twocolumn,notitlepage,amsmath,amstex,amssymb,citeautoscript,longbibliography]{revtex4-2}

\usepackage{amsmath,amssymb,amsfonts}
\usepackage{graphicx}
\usepackage{tikz}
\usepackage{physics}
\usepackage{amsthm}
\usepackage{mathtools}
\usepackage{hyperref}
\usepackage{algorithm,algpseudocode}
\usepackage{bm}
\usepackage{bbm}
\usepackage{multirow}
\usepackage{booktabs}
\usepackage[caption=false]{subfig} 
\DeclareMathOperator*{\argmax}{arg\,max}

\DeclarePairedDelimiter\floor{\lfloor}{\rfloor}

\newcommand{\ER}{Erd\"{o}s-R\'{e}nyi }
\newcommand{\vgamma}{\bm{\gamma}}
\newcommand{\vbeta}{\bm{\beta}}

\newcommand{\Opt}[1]{(\gamma^*_{#1},\beta^*_{#1})}
\newcommand{\Gbvec}{(\vgamma,\vbeta)}

\newcommand{\Gb}[1]{(\gamma_{#1},\beta_{#1})}

\newcommand{\itlw}[2]{\text{ITLW}(#1, #2)}
\newcommand{\BigO}{\mathcal{O}}

\newtheorem{definition}{Definition}

\newtheorem{corollary}{Corollary}

\begin{document}
\title{Iterative Layerwise Training for Quantum Approximate Optimization Algorithm}

\author{Xinwei Lee}
\email{xwlee@cavelab.cs.tsukuba.ac.jp}
\author{Xinjian Yan}
\author{Ningyi Xie}
\author{Dongsheng Cai}
\affiliation{University of Tsukuba, Ibaraki Prefecture, Japan}
\author{Yoshiyuki Saito}
\author{Nobuyoshi Asai}
\affiliation{University of Aizu, Fukushima Prefecture, Japan}

\date{\today}

\begin{abstract}
    The capability of the quantum approximate optimization algorithm (QAOA) in solving the combinatorial optimization problems has been intensively studied in recent years due to its application in the 
    quantum-classical hybrid regime. Despite having difficulties that are innate in the variational quantum algorithms (VQA), such as barren plateaus and the local minima problem, QAOA remains one of the
    applications that is suitable for the recent noisy intermediate scale quantum (NISQ) devices. Recent works have shown that the performance of QAOA largely depends on the initial parameters, which motivate
    parameter initialization strategies to obtain good initial points for the optimization of QAOA. On the other hand, optimization strategies focus on the optimization part of QAOA instead of the parameter initialization.
    Instead of having absolute advantages, these strategies usually impose trade-offs to the performance of the optimization problems. One of such examples is the 
    layerwise optimization strategy, in which the QAOA parameters are optimized layer-by-layer instead of the full optimization. The layerwise strategy costs less in total compared to the full optimization, in exchange of lower 
    approximation ratio. In this work, we propose the iterative layerwise optimization strategy and explore the possibility for the reduction of optimization cost in solving problems with QAOA. Using numerical simulations,
    we found out that by combining the iterative layerwise with proper initialization strategies, the optimization cost can be significantly reduced in exchange for a minor reduction in the approximation ratio.
    We also show that in some cases, the approximation ratio given by the iterative layerwise strategy is even higher than that given by the full optimization.
\end{abstract}

\maketitle

\section{Introduction}
The quantum approximate optimization algorithm (QAOA) is a quantum-classical hybrid algorithm introduced in 2014~\cite{farhi2014quantum}, aimed to solve combinatorial optimization problems
on a universal quantum computer using the paradigm of the discretized adiabatic quantum computation~\cite{farhi:qaa}.
Many recent works have explored the potential of QAOA to have quantum advantage in some of the optimization problems against their classical
counterparts~\cite{lloyd2018quantum,Crooks2018PerformanceOT,leo2020,Moussa_2020,Marwaha_2021},
and many variants of QAOA has been suggested to satisfy their respective needs~\cite{new-qaoa,adapt-qaoa,grover-mixer,ma-qaoa}.
However, QAOA inherits properties that are innate in the variational quantum algorithms (VQA), such as the existence of the barren plateaus in the optimization landscape~\cite{Grant_2019,barren_vqa2021,Cerezo_2021,Wang_2021},
and the exponential increase of local minima as the circuit depth increases~\cite{Guerreschi_2019,sack-greedy2023}, rendering the optimization of QAOA difficult. 
To tackle these problems, various initialization strategies are proposed so that QAOA can be initialized near the desired optima~\cite{leo2020,sack2021quantum,lee2021parameters,multistart},
aiding the convergence of the classical optimizers. 

Besides the initialization strategies, the optimization strategies are also essential in improving the optimization of QAOA.
The layerwise strategy (or layerwise training)~\cite{campos2021} is an optimization strategy which trains the QAOA parameters layer-by-layer.
In the layerwise training, only the latest parameters are updated while the other parameters are fixed constant. 
This has led to a reduction in the optimization cost, but also causes the optimization to occur in a restricted search space, 
and the premature saturation (saturation before the approximation ratio reaches 1) of the layerwise training is reported in~\cite{campos2021} for a projector Hamiltonian.
We found that the premature saturation also occurs in the Max-cut ($ZZ$) Hamiltonian.
However, the layerwise training is shown to perform better in the quantum neural network~\cite{lw-qnn}.

In this work, we propose the iterative layerwise (abbreviated as ITLW) strategy which improves the layerwise strategy and helps to escape the premature saturation.
We found out that the trainability of the parameters in ITLW is highly sensitive to the initial parameters used.
Therefore, by combining ITLW with some initialization strategies, e.g. the bilinear strategy and the TQA initialization, we successfully reduce the optimization cost (evaluated by the total number of function evaluations)
significantly in exchange of a minor reduction in the approximation ratio. 
For some small number of iterations $k=2$, ITLW managed to reduce the cost by almost half with a $4\times 10^{-3}$ reduction of the approximation ratio.

The structure of the paper is summarized as follows. In Sec.~\ref{sec:background}, we introduce the background of QAOA and the layerwise training.
In Sec.~\ref{sec:itlw}, we discuss the idea of the ITLW strategy and how the expected optimization cost scales with the QAOA circuit depth.
In Sec.~\ref{sec:results}, we present and discuss the simulation results for ITLW combined with the initialization strategies (bilinear and TQA).
Lastly, we conclude in Sec.~\ref{sec:conclusion}.

\section{QAOA and the Layerwise Training}\label{sec:background}
The objective of QAOA is to maximize the expectation of some cost Hamiltonian $H_z$ with respect to the ansatz state $\ket{\psi\Gbvec}$ prepared by the evolution of the alternating operators:
\begin{equation}
	\ket{\psi_p\Gbvec} = \prod^p_{j=1} e^{-i\beta_j H_x}e^{-i\gamma_j H_z}\ket{+}^{\otimes n} \label{eqn:ansatz}.
\end{equation}
where $\vgamma = (\gamma_1, \gamma_2, \ldots, \gamma_p)$ and $\vbeta = (\beta_1, \beta_2, \ldots, \beta_p)$ are the $2p$ variational parameters.
$\ket{+}^{\bigotimes n}$ corresponds to $n$ qubits in the ground state of $H_x = \sum_{j=1}^n X_j$, where $X_j$ is the Pauli $X$ operator acting on the $j$-th qubit.

The symbol $p$ is commonly used to denote the circuit depth of QAOA. It is defined as how many pairs of operators (layers) that is in the quantum circuit.
In our context, we use the term \emph{layer}, or usually $l$, to represent which layer of the circuit that is concerned. 
The total number of layers is equal to the circuit depth.
We also use the subscript $p$ to represent a quantity at the circuit depth $p$, e.g., $\ket{\psi_p}$ means the ansatz state $\ket{\psi}$ at circuit depth $p$.

\begin{definition}[Depth of QAOA circuit]
    The depth of a QAOA circuit is defined the number of alternating operator pairs (the cost and mixer Hamiltonians) applied to the initial state, usually denoted as $p$.
    A system of the standard QAOA consists of $2p$ parameters.
\end{definition}

\begin{definition}[Layer of QAOA circuit]
    The layer of a QAOA circuit is defined as the index of the depth, starting from 1: $l=1,2,...,p$. It is used to represent which layer of the circuit is concerned.
\end{definition}

In this paper, we consider the infamous Max-cut problem, which aims to divide a graph into two parts, with the maximum number of edges between them. 
The cost Hamiltonian $H_z$ for the Max-cut problem for an unweighted graph $G = (V,E)$ is given as
\begin{equation}
	H_z = \frac{1}{2}\sum_{(j, k)\in E}(\mathbbm{1} - Z_j Z_k),
\end{equation}
where $Z_j$ is the Pauli $Z$ operator acting on the $j$-th qubit.
We define the expectation of $H_z$ with respect to the ansatz state in Eq.~(\ref{eqn:ansatz}):
\begin{equation}
	F(\vgamma,\vbeta) = \expval{H_z}{\psi(\vgamma,\vbeta)}.
	\label{eqn:fp}
\end{equation}
Solving the problem with QAOA is equivalent to maximizing Eq.~(\ref{eqn:fp}), with respect to the variational parameters $\vgamma$ and $\vbeta$. This can be done by
a classical optimizer which search for the maximum $F$ and the parameters that maximize it:
\begin{equation}
	(\vgamma^*,\vbeta^*) = \argmax_{\vgamma,\vbeta} F(\vgamma,\vbeta),
\end{equation}
where the superscript * denotes quasi-optimal parameters returned by the classical optimizer.
We also define the \emph{approximation ratio} $\alpha$ as
\begin{equation}
	\alpha = \frac{F(\vgamma^*,\vbeta^*)}{C_{\text{max}}},
	\label{eqn:alpha}
\end{equation}
where $C_{\text{max}}$ is the maximum cut value for the graph. The approximation ratio is a typical evaluation metric indicating how near the solution given by QAOA is to the true solution, $0\leq\alpha\leq1$, with
the value of 1 nearer to the true solution. 

Theoretically, $\alpha$ approaches 1 as $p\rightarrow\infty$. However, due to the exponential increase of the number of local minima at larger $p$, optimizers tend to converge to an undesired local minima if QAOA is 
initialized randomly. This led to various initialization strategies being proposed to obtain good quality solutions for QAOA. 
Most of the initialization strategies exploit the fact that the optimal parameters of QAOA exhibit a smooth and monotonous pattern, with monotonically increasing $\gamma_i$ and monotonically decreasing $\beta_i$, 
within the QAOA periodic bound.
Depth-progressive initialization strategies, such as the Bilinear strategy~\cite{bilinear} and INTERP~\cite{leo2020}, are known to produce high $\alpha$ by fine tuning the parameters depth-by-depth 
so that the errors do not accumulate.
This is at the cost of depth-progressive strategies requiring the optimization of every $p'$ for $p'=1, 2, ..., p$, which is very costly.
Strategy like TQA~\cite{sack2021quantum} uses direct initialization at $p$ without the requirement of starting from small depths, but the result is also sub-optimal at large depths~\cite{sack-greedy2023}.

Another type of strategy is the optimization strategy, in which the manipulation of parameters occurs during the optimization step. The layerwise strategy~\cite{campos2021} falls under this category. 
The layerwise strategy only updates the parameters of the latest depth.
This form of training method greatly reduces the optimization (classical computational) cost, as only 2 variables need to be optimized, compared with the full optimization of $2p$ variables.
However, the layerwise strategy encounters premature saturation in the approximation ratio $\alpha$, due to the restricted search space with only 2 latest variables.
This behavior of premature saturation is observed in the projector Hamiltonian~\cite{campos2021} and the Max-cut Hamiltonian~\cite{bilinear}.
Here, we want to make a clear distinction between the initialization strategies and the optimization strategies as two different categories,
for both of them can be combined to work together, or they can work independently on their own.

\section{Iterative Layerwise}\label{sec:itlw}
We propose the iterative layerwise (ITLW) strategy which iteratively updates the QAOA parameters in the layerwise manner, for a certain number of iterations $k$.
We use the notation $\itlw{k}{p}$ to denote the ITLW strategy with $k$ iterations for circuit depth $p$.
We can say that the layerwise training is similar to a special case of ITLW at $k=1$, i.e, layerwise for circuit depth $p$ is similar to $\itlw{1}{p}$.
However, there is a subtle difference between the layerwise training discussed in~\cite{campos2021} and $\itlw{1}{p}$. 
In layerwise, the parameters are optimized on a shallower circuit starting from $p=1$, then the QAOA layers are gradually appended as the depth increases.
For $\itlw{1}{p}$, we first generate $2p$ random parameters for a desired depth $p$, then the parameters for each layer, $\Gb{l}$ for $l=1,2,...,p$, are optimized separately.
This modification is done for a simpler repetition procedure at larger $k$. Table~\ref{tab:lw-diff} shows an example of both strategies for the target depth $p=3$.

\begin{table}[t]
    \caption{Difference between the layerwise strategy and $\itlw{1}{p}$. Layerwise starts with shallower circuits, then more layers are appended as the depth increases. %
    For ITLW, the random full parameter vector of a desired depth $p$ is generated, then they are optimized layer-by-layer.}
    \centering
    \renewcommand{\arraystretch}{1.5} 
    \begin{tabular}{ccccc} 
        \toprule
        \multirow{2}{1cm}{Layer} & \multicolumn{2}{c}{Layerwise} & \multicolumn{2}{c}{$\itlw{1}{p}$} \\
        \cline{2-5}
         & Optimized & Fixed & Optimized & Fixed \\
        \midrule
        $l=1$ & $\Gb{1}$ & -- & $\Gb{1}$ & $(\gamma_2, \gamma_3, \beta_2, \beta_3)$ \\
        $l=2$ & $\Gb{2}$ & $(\gamma_1, \beta_1)$ & $\Gb{2}$ & $(\gamma_1, \gamma_3, \beta_1, \beta_3)$ \\
        $l=3$ & $\Gb{3}$ & $(\gamma_1, \beta_1, \gamma_2, \beta_2)$ & $\Gb{3}$ & $(\gamma_1, \gamma_2, \beta_1, \beta_2)$ \\
        \bottomrule    
    \end{tabular}
    \renewcommand{\arraystretch}{1}
    \label{tab:lw-diff}
\end{table}

A brief description for the iterative layerwise strategy is as follows. First, choose a QAOA circuit depth $p$ and the number of iterations $k$. 
Then, initialize a parameter buffer $\bm{\Phi}$ with random parameters of length $2p$: $\bm{\Phi} := (\gamma_1, ..., \gamma_p, \beta_1, ..., \beta_p)$.
For each iteration, the parameters are optimized in a layerwise manner:
each pair of parameters $\Gb{l}$ are optimized layer-by-layer and update $\bm{\Phi}$ with the optimized parameters $\Opt{l}$, for $l=1,2,...,p$.
This procedure is repeated for $k$ iterations. Note that the parameter buffer $\bm{\Phi}$ is inherited to the next iteration.
The algorithm for $\itlw{k}{p}$ is summarized in Algorithm~\ref{alg:itlw}.

\begin{algorithm}[H]
    \caption{Iterative Layerwise}
    \textbf{Input:} No. of iterations $k$, circuit depth $p$.
    \begin{algorithmic}[1]    
        \State Initialize parameter buffer $\bm{\Phi} := (\gamma_1, ..., \gamma_p, \beta_1, ..., \beta_p)$.
        \For{$k'=1...k$}
            \For{$l=1...p$}
                \State $\Opt{l} := \argmax_{\gamma_l, \beta_l} F(\bm{\Phi})$
                \State Update $\bm{\Phi}$ with $\Opt{l}$.
            \EndFor
        \EndFor
        \State $\bm{\Phi^*_p} := \bm{\Phi}$
    \end{algorithmic}
    \textbf{Output:} Optimized parameters $\bm{\Phi}^*_p$ and its expectation $F_p(\bm{\Phi}^*)$.
    \label{alg:itlw}
\end{algorithm}

\begin{figure}
    \centering
    \includegraphics[width=\columnwidth]{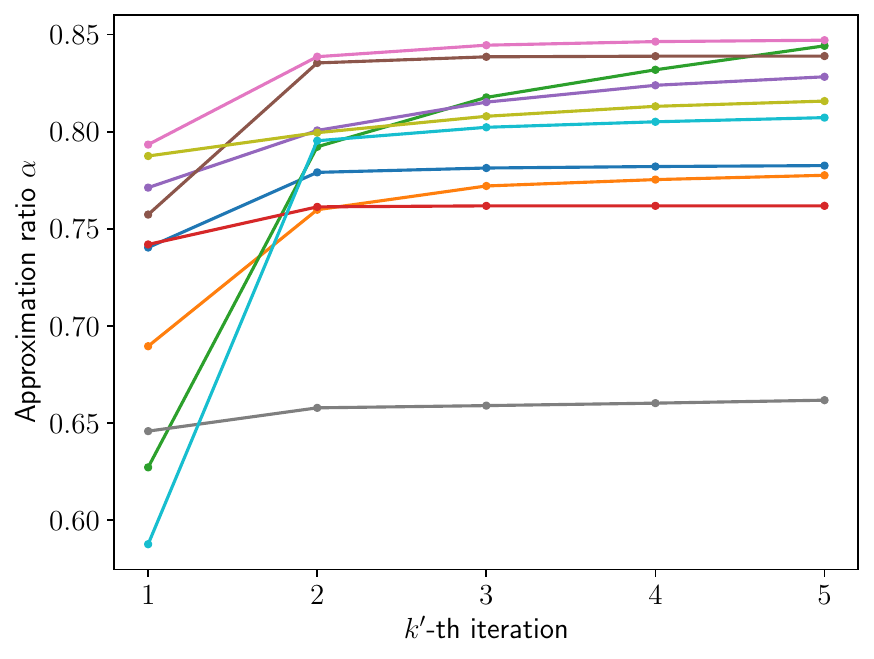}
    \caption{$\itlw{5}{5}$ applied on solving the QAOA of a 6 vertices graph. The lines represent 10 different sets of initial parameters starting at $k'=1$. The trainability varies depending on the initial parameters. %
    Some $\alpha$'s can be improved through layerwise iterations, while some of them stay almost the same throughout the iterations.}
    \label{fig:random-init}
\end{figure}

Figure~\ref{fig:random-init} shows the variation of the approximation ratio $\alpha$ with the iteration index $k'$ of $\itlw{5}{5}$, for a graph with 6 vertices, for 10 random sets of initial parameters.
Each line represents the variation for a set of initial parameters. 
It can be seen that the trainability (whether $\alpha$'s improve through the iterations) depends strongly on the initial parameters used to initialize ITLW.
Some of the parameters improves greatly through layerwise iterations, while some barely increase after the iterations.
This, again, leads to the requirement for a good selection of initial parameters for ITLW.

\subsection{Combining with initialization strategies}
We expect a better performance for ITLW using good initializations. 
In order to generate better initial parameters, we use two types of initialization strategies along with ITLW: the bilinear strategy~\cite{bilinear} (depth-progressive) and the TQA strategy~\cite{sack2021quantum} 
(direct initialization). 
The details of the strategies are described in Appendix~\ref{sec:detail-init}, but we will introduce them briefly here.

The bilinear strategy is a depth-progressive strategy which generates the initial parameters for the current depth $p$ with the optimized parameters of the two previous depths: $p-1$ and $p-2$.
The strategy is motivated by the linear-like pattern (a smooth, monotonous variation) that QAOA parameters exhibits in two directions: the parameter index and the circuit depth. 
The optimal parameters in $p-1$ and $p-2$ are extrapolated by taking linear differences to generate the new initial parameters for $p$. 
Therefore, optimization is required at every depth in order to find their respective optimal parameters.

The TQA strategy offers a more direct way to initialize QAOA, without requiring the information of the parameters from previous depths. 
TQA starts by establishing a linear relation between the parameters ($\gamma_i$ and $\beta_i$) and the total annealing time $T$.
Then, the gradient $T$, of the straight line passing through the origin, is optimized to find the straight line which gives the minimum/maximum expectation.
QAOA is then initialized with the straight line with the optimized gradient $T^*$.

The procedure of ITLW combined with initialization strategies is essentially the same as in Algorithm~\ref{alg:itlw}. 
The initialization step is replaced with their respective initialization procedures instead of random initialization.
For depth-progressive strategies, an outer loop of $p'=1,2,...,p$ is added.

\subsection{Optimization cost}\label{sec:opt-cost}
There are many evaluation metrics for the optimization cost of a classical optimizer.
We use the number of function evaluations (number of function calls to the quantum circuit) as our evaluation metric for the optimization cost,
as this quantity usually scales with the number of parameters to be optimized.
Also, in classical simulations, the optimization time is mostly determined by the number of function evaluations.

Assume that the number of function evaluations scales with the number of parameters, and hence the circuit depth $p$. 
Let $V(p)$ be the number of evaluations required to optimize the expectation function $F_p$ given in Eq.~(\ref{eqn:fp}), for the entire parameter space with $2p$ parameters.
We call this the full optimization.
For a certain fixed $p$, it requires $V(p)$ evaluations for full optimization, while it requires $kpV(1)$ evaluations for $\itlw{k}{p}$.
This is because for $\itlw{k}{p}$, 2 parameters are optimized per layer for a total of $p$ layers, and for $k$ iterations.
Therefore, it is expected that at large $p$, ITLW will have an advantage in cost against the full optimization if $V(p)$ scales superlinearly with $p$, 
i.e., $V(p) = \BigO(p^m)$ for $m>1$, as $kpV(1) = \BigO(p)$, given that $k$ is a sufficiently small constant.

When used with depth-progressive initialization strategies~\cite{bilinear,leo2020}, full optimization requires a total of $S(p) \equiv \sum_{p'=1}^p V(p')$ evaluations.
This leads to a total cost of $S(p) = \BigO(p^{m+1})$, given that $V(p) = \BigO(p^m)$.
On the other hand, ITLW calls for a total of $\sum_{p'=1}^p kp'V(1) = \BigO(p^2)$ evaluations.
This shows ITLW is also polynomially faster than full optimization at large $p$ when combined with depth-progressive initialization, under the assumption that $V(p)$ is superlinear.

\begin{corollary}\label{cor:vp}
    At large circuit depth $p$, ITLW is polynomially faster than full optimization if $V(p)$ is superlinear w.r.t. $p$ ($V(p) = \BigO(p^m)$ for $m>1$).
\end{corollary}

\begin{figure*}[t]
    \subfloat[]{
        \begin{minipage}[c]{0.32\textwidth}
            \centering
            \includegraphics[width=\textwidth]{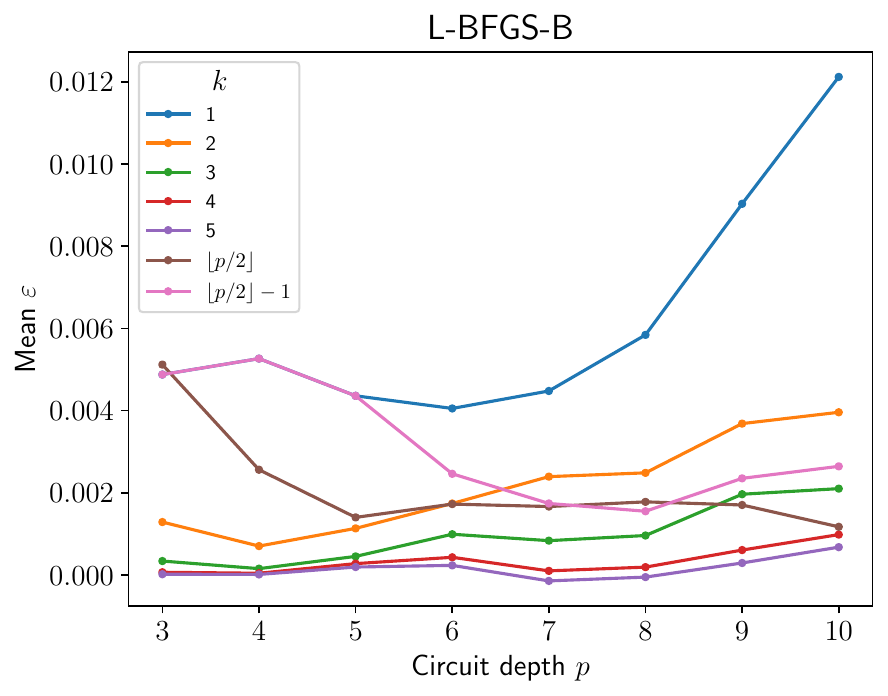}
        \end{minipage}
    }
    \hfill
    \subfloat[]{
        \begin{minipage}[c]{0.32\textwidth}
            \centering
            \includegraphics[width=\textwidth]{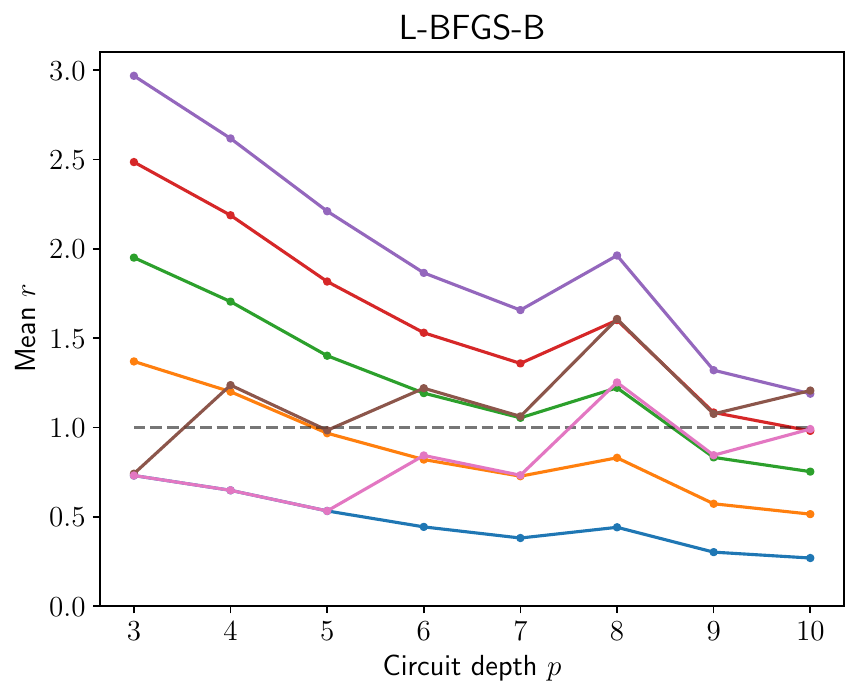}
        \end{minipage}
    }
    \hfill
    \subfloat[]{
        \begin{minipage}[c]{0.32\textwidth}
            \centering
            \includegraphics[width=\textwidth]{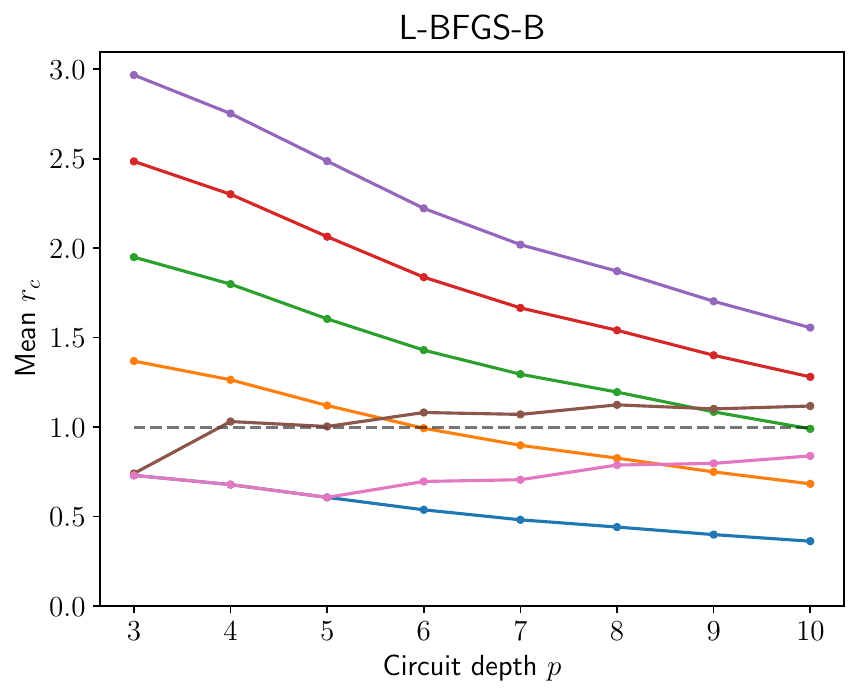}
        \end{minipage}
    }
    \hfill
    \subfloat[]{
        \begin{minipage}[c]{0.32\textwidth}
            \centering
            \includegraphics[width=\textwidth]{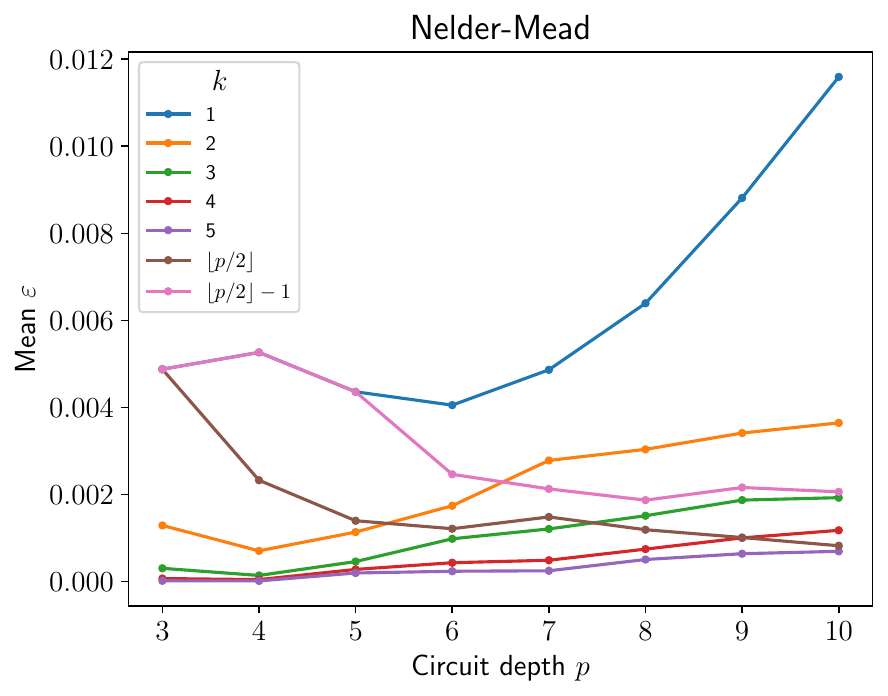}
        \end{minipage}
    }
    \hfill
    \subfloat[]{
        \begin{minipage}[c]{0.32\textwidth}
            \centering
            \includegraphics[width=\textwidth]{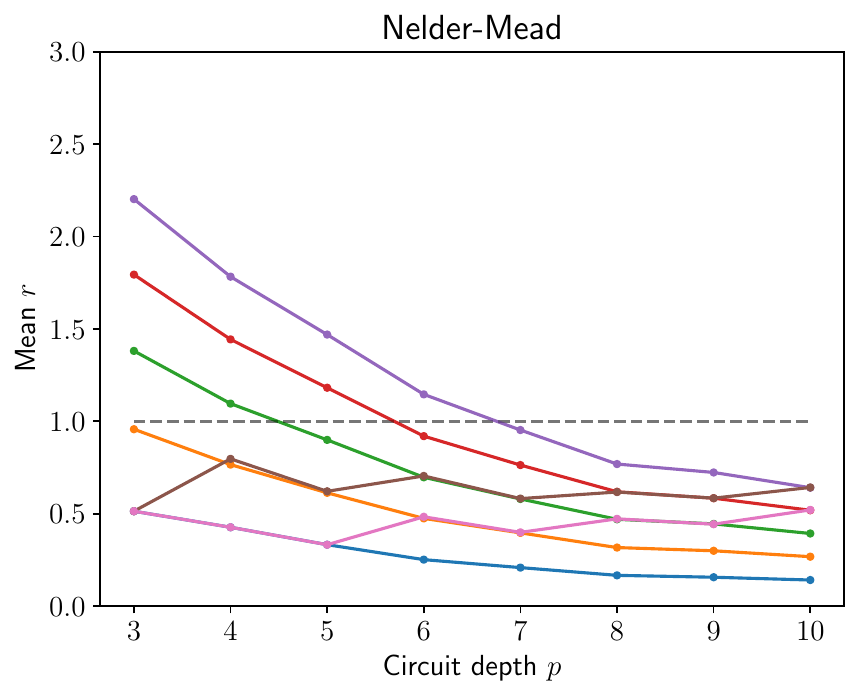}
        \end{minipage}
    }
    \hfill
    \subfloat[]{
        \begin{minipage}[c]{0.32\textwidth}
            \centering
            \includegraphics[width=\textwidth]{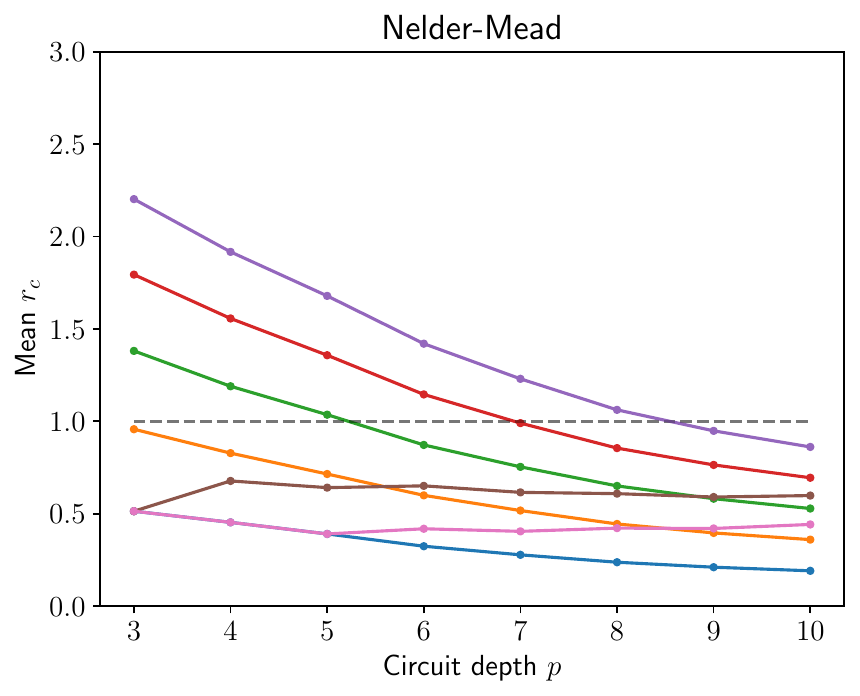}
        \end{minipage}
    }
    \caption{The results for ITLW with bilinear initialization. The columns show the mean error in approximation ratio $\varepsilon$, the cost reduction ratio $r$, and the cumulative-cost reduction ratio $r_c$, respectively. %
    Different color of lines show the results for different number of iterations $k$. %
    The top row shows the results for the L-BFGS-B optimizer. The bottom row shows the results for the Nelder-Mead optimizer. %
    The dashed gray lines in the $r$ and $r_c$ figures show the critical ratio of 1 when ITLW has advantage over the full optimization.}
    \label{fig:bl-results}
\end{figure*}

\begin{figure*}[t]
    \subfloat[]{
        \begin{minipage}[c]{0.45\textwidth}
            \centering
            \includegraphics[width=\textwidth]{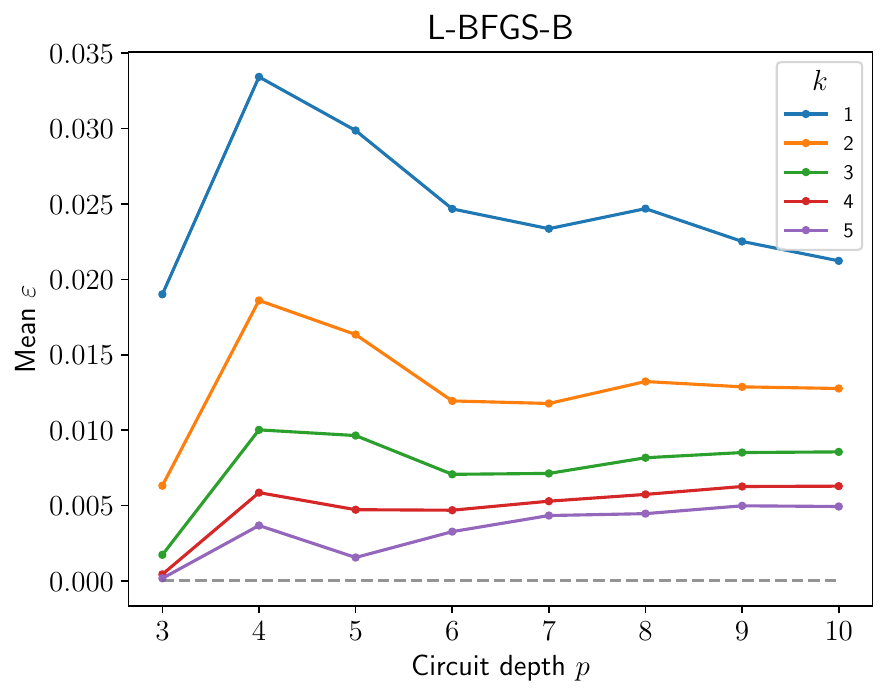}
        \end{minipage}
    }
    \hfill
    \subfloat[]{
        \begin{minipage}[c]{0.45\textwidth}
            \centering
            \includegraphics[width=\textwidth]{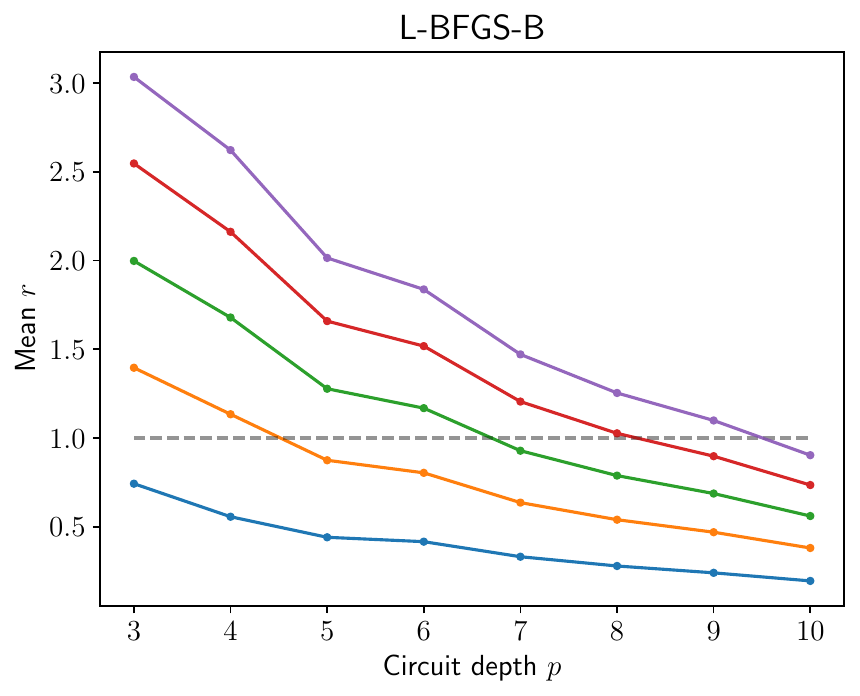}
        \end{minipage}
    }
    \hfill
    \subfloat[]{
        \begin{minipage}[c]{0.45\textwidth}
            \centering
            \includegraphics[width=\textwidth]{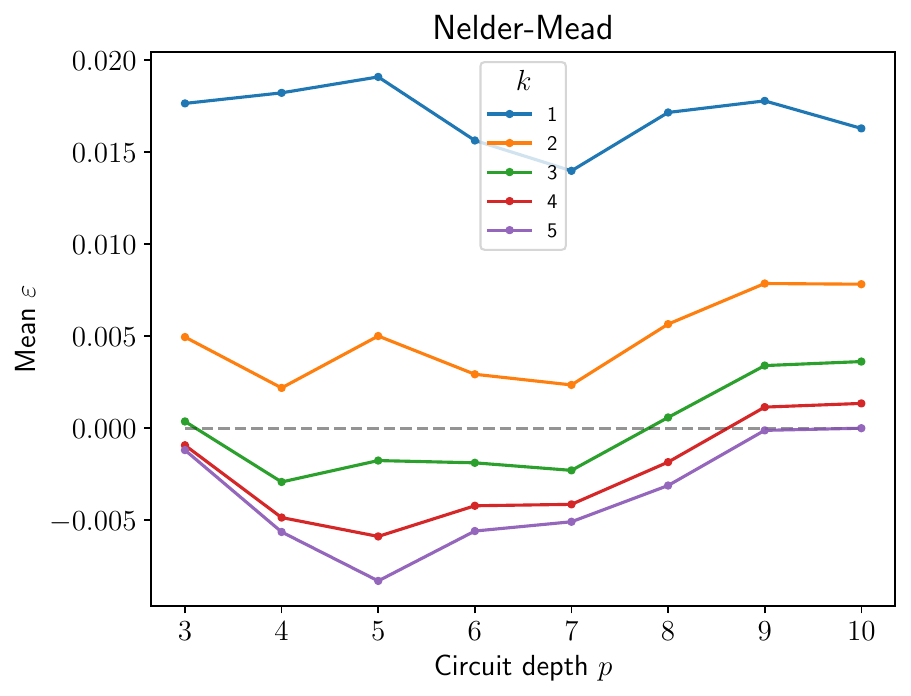}
        \end{minipage}
    }
    \hfill
    \subfloat[]{
        \begin{minipage}[c]{0.45\textwidth}
            \centering
            \includegraphics[width=\textwidth]{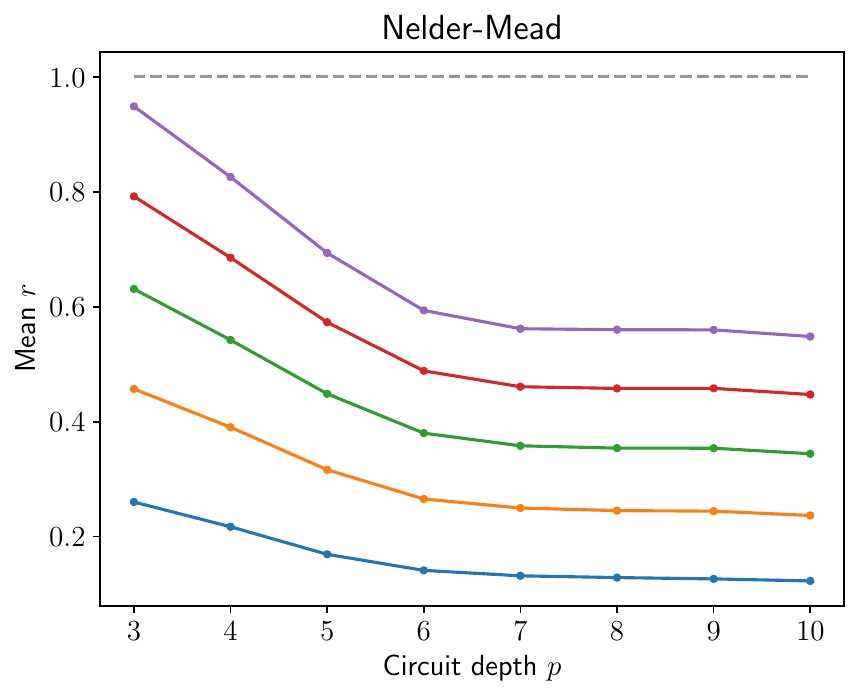}
        \end{minipage}
    }
    \caption{The results for ITLW with TQA initialization. The mean error in approximation ratio $\varepsilon$ and the cost reduction ratio $r$, respectively. %
    The top row shows the results for the L-BFGS-B optimizer and the bottom row shows the results for the Nelder-Mead optimizer. %
    The dashed gray lines show the critical values $\varepsilon=0$ and $r=1$ when ITLW has advantage over the full optimization.}
    \label{fig:tqa-results}
\end{figure*}

\section{Results and Discussions}\label{sec:results}
We apply the iterative layerwise strategy $\itlw{k}{p}$ with different initialization strategies in solving the Max-cut problem using QAOA.
Our experimental settings are as follows:
\begin{itemize}
    \item The quantum circuits are simulated using Qiskit statevector simulation (exact simulation of quantum circuits).
    \item We use 30 non-isomorphic graph instances with number of vertices $n=10$ to $n=12$. The graph instances are generated randomly, 
    comprising different classes including the 3- and 4-regular graphs and the \ER graphs with varied edge probabilities.    
    \item Two different initialization strategies are used: the bilinear strategy~\cite{bilinear} and the TQA strategy~\cite{sack2021quantum}.
    \item The parameters are bounded by $\vgamma\in [0, \pi)^p$ and $\vbeta\in [0, \pi/2)^p$ due to the symmetry and the periodicity of QAOA operators~\cite{bilinear,leo2020,Lotshaw_2021,sack-greedy2023}, 
    so that the optimal parameters reproduce the linear-like pattern for good initialization.
    \item The performances are compared with two different classical optimizers that are widely used in other works: the Limited-memory Broyden–Fletcher–Goldfarb–Shanno Bounded (L-BFGS-B) algorithm~\cite{l-bfgs-b}
    and the Nelder-Mead algorithm~\cite{neldermead}. L-BFGS-B is gradient-based and Nelder-Mead is gradient-free. The hyperparameters for the optimizers are the default values provided by the SciPy package.
    \item The performance of the strategy is evaluated in terms of two different metrics: the approximation ratio $\alpha$ (for the quality of the solution) and the number of function evaluations (for the optimization cost).    
    \item The performances of different number of iterations for ITLW are compared: $k=1,2,3,4,5$. We also consider adaptive values for $k$ that depends on the circuit depth $p$, i.e., $k=\floor{p/2}$ and $k=\floor{p/2}-1$,
    where $\floor{\cdot}$ represents the floor function.
\end{itemize}

The main comparison is done between the results produced by ITLW and the full optimization (FO). 
We define the \emph{error in the approximation ratio} $\varepsilon$ as
\begin{equation}
    \varepsilon = \alpha_{\text{FO}} - \alpha_{\text{ITLW}},
    \label{eqn:error}
\end{equation}
where $\alpha_{\text{FO}}$ is the approximation ratio obtained from FO, and $\alpha_{\text{ITLW}}$ is the approximation ratio from ITLW.
Note that to write Eq.~(\ref{eqn:error}) out explicitly, it will become
\begin{equation}
    \varepsilon_p = \alpha_{p,\text{FO}} - \alpha_{p,\text{ITLW}}.
\end{equation}
However, for simplicity, we will omit the subscript $p$ as we assume that the readers are aware that this quantity (and those defined later) are different at different $p$'s.
The quantity $\varepsilon$ allows us to know how much the $\alpha$ given by ITLW deviates from that of FO. 
If $\varepsilon > 0$, FO approximates better, and if $\varepsilon < 0$, ITLW approximates better.
Note that since $0 \leq \alpha \leq 1$, $\varepsilon$ can take values between the range $-1 \leq \varepsilon \leq 1$.

We define another quantity, known as the \emph{cost reduction ratio} $r$, to evaluate how much reduction in the number of function evaluations for ITLW:
\begin{equation}
    r = \frac{\tilde{V}_{\text{ITLW}}}{\tilde{V}_{\text{FO}}}.
\end{equation}
We use the notation $\tilde{V}$ to denote the empirical number of function evaluations (instead of the theoretical $V$ in the previous section). 
If $r < 1$, it means there is reduction in the number of function evaluations for ITLW compared to FO.
If $r > 1$, it means ITLW costs more than FO. 

For depth-progressive strategies, we introduce another quantity called the \emph{cumulative-cost reduction ratio} $r_c$:
\begin{equation}
    r_c = \frac{\sum_p \tilde{V}_{p,\text{ITLW}}}{\sum_p \tilde{V}_{p,\text{FO}}},
\end{equation}
where $\sum_p \tilde{V}_p$ means the sum of the number of function evaluations up to the target depth $p$. 
This is a more objective quantity for the total cost as we need to consider the costs for every depths for depth-progressive strategies.

When calculating $\varepsilon$, $r$, and $r_c$, we use the values that is generated under the same experimental conditions, e.g., same optimizer, same circuit depth, and same initialization strategy,
so that the only difference between the values is caused by the different optimization strategies, i.e., ITLW and FO.
However, since FO does not have the parameter $k$, we use the same value for all $k$'s when comparing with ITLW, while keeping the other conditions the same.
For a simpler interpretation, having lower values of the quantities $\varepsilon$ and $r$ (and $r_c$) implies better results.

\subsection{Bilinear initialization}
Fig.~\ref{fig:bl-results} shows the results of ITLW with bilinear initialization, averaged over the 30 graphs that we considered.
For the bilinear initialization, we initialize from $p=3$, then slowly raising the depth to $p=10$. 
Each depth is optimized using the ITLW strategy. 
Fig.~\ref{fig:bl-results}(a), (b) and (c) shows the results for the L-BFGS-B optimizer; (d), (e) and (f) shows the results for the Nelder-Mead optimizer.
It can be observed that the overall trend does not change much when comparing (a) and (d), implying that the choice of optimizers does not have much effect on the errors in $\alpha$ under this experimental condition.
Fig.~\ref{fig:bl-results}(a) shows the errors $\varepsilon$ decreases as the number of iterations $k$ increases. 
This tells that the parameters generated from the bilinear strategy is trainable as the error decreases with increasing $k$ ($\alpha_{\text{FO}}$ is the same for all $k$ for a certain $p$).
Also, the strategy is able to achieve a considerably small error $\varepsilon < 4\times 10^{-3}$, even for a small $k=2$, with decreasing $\varepsilon$ as $k$ increases.

Fig.~\ref{fig:bl-results}(b) and (e) shows the cost reduction ratio $r$ averaged over 30 graph instances. 
The gray dashed lines show the critical value $r=1$, in which ITLW costs less (faster) than FO if $r$ is below the critical value.
Overall, the Nelder-Mead optimizer has lower $r$ compared to the L-BFGS-B optimizer. 
The general trend shows that $r$ is lower for larger circuit depth $p$, caused by the difference between the costs of ITLW and FO mentioned in Sec.~\ref{sec:opt-cost}.
This result agrees with Corollary~\ref{cor:vp}.
Recall that the costs for ITLW and FO differ by a polynomial degree of $p$, so we would expect that more cost is saved at larger $p$.
For L-BFGS-B, at $p=10$, $\itlw{4}{10}$ is able to achieve a borderline advantage at $r\approx 1$. Although $\itlw{5}{10}$ fails to achieve an advantage, it is expected to have an even lower $r$ if $p$ is further increased.
For Nelder-Mead, the cost reduction ratio is lower than 1 for all $k$ at the target depth $p=10$. 
Despite the cost is lower at larger circuit depths, the errors has a tendency to increase at larger $p$ as shown in Fig.~\ref{fig:bl-results}(a) and (d).
Thus, the ITLW strategy is more like a trade-off strategy rather than a strategy with an absolute advantage. 
The approximation ratio is compromised in exchange for a reduction in cost. 
However, we still need to consider whether it is worth to exchange a reduction in $\alpha$ by $4\times 10^{-3}$ given that the cost is almost halved (for the case of $k=2$).

For the adaptive number of iterations $k=\floor{p/2}$, the mean error at $p=10$ is near to that of $k=4$ for L-BFGS-B, and is near to the mean error of $k=5$ for Nelder-Mead.
It can be observed from Fig.~\ref{fig:bl-results}(c) and (f) that the total cost for the ITLW is less than that of FO for this adaptive $k$ (comparing the red and brown lines),
which means it is worth switching to the adaptive strategy.
However, since the cost for the constant $k$ will decrease further for larger $p$, we expect that the adaptive $k$ will cost more than the constant $k$ at larger $p$.
For $k=\floor{p/2} - 1$, the same patterns occur. The mean error at $p=10$ is near to that of $k=3$ and the cost is less. The cost is expected to grow for larger $p$.
It is interesting to compare both adaptive values for the L-BFGS-B optimizer, where the total cost of ITLW is less than that of FO for $\floor{p/2} - 1$, but is more than FO for $\floor{p/2}$,
despite only having 1 iteration difference for each depth.

\subsection{TQA initialization}
Fig.~\ref{fig:tqa-results} shows the results for ITLW with the TQA initialization, relative to the full optimization. 
The general trend for $\varepsilon$ (shown in Fig.~\ref{fig:tqa-results}(a) and (c)) with increasing $k$ remains as expected.
As $k$ increases, the overall mean $\varepsilon$ decreases, which shows the approximation ratio $\alpha$ improves as we increase the number of iteration. 
The Nelder-Mead optimizer also gives less error than the L-BFGS-B optimizer for the same $k$. 
Also, for the Nelder-Mead optimizer, as $k$ increases, we can see that ITLW yields higher $\alpha$, as much as $8\times 10^{-3}$ higher than FO at some of the circuit depths,
which is rarely seen in the ITLW with bilinear initialization.
This might be caused by the $\alpha$ obtained from TQA is relatively lower than that obtained from bilinear, giving more room of improvement when ITLW is applied. 

Fig.~\ref{fig:tqa-results}(b) and (d) shows the respective cost reduction ratio for L-BFGS-B and Nelder-Mead.
For both of the optimizers, the mean $r$ decreases monotonically as $p$ increases.
Comparing the cost reduction ratio of the two optimizers, we can see a large difference in the magnitude of the values of $r$. 
However, for L-BFGS-B, the costs still pass through the critical value $r=1$ as $p$ increases. 
For Nelder-Mead, $r<1$ for all $k$'s, despite having higher $\alpha$ than FO for $k=3,4,5$ at some of the depths.
This shows that the Nelder-Mead optimizer exhibits significant cost advantage with TQA initialization.

Since TQA is a direct initialization method, we did not include the results ($\varepsilon$ and $r$) for the adaptive $k$ as they are just the same as those at their respective number of iterations.

\section{Conclusions}\label{sec:conclusion}
Our research brought out the possibility of combining initialization strategies and optimization strategies in QAOA.
We proposed the iterative layerwise (ITLW) optimization strategy for QAOA and use it to solve the Max-cut problems with the size of 10 to 12 qubits.
As an improvement to the layerwise training, ITLW prevents the premature saturation that occurs in the layerwise training, which allows the parameters to be further trained to achieve a higher approximation ratio.
However, the trainability of the parameters is still highly sensitive to the initial parameters, so we combine ITLW with the bilinear and TQA initialization strategies for a better initialization.
Instead of being a strategy with an absolute advantage, the strategy imposes a trade-off between the approximation ratio and the optimization cost.
The simulation results show that ITLW has lower optimization cost in exchange for a lower approximation ratio on average, compared to the full optimization (FO).
The cost reduction gets more significant as the QAOA circuit depth increases.
For the problem instances that we considered, ITLW is able to reduce the cost by half in exchange for a reduction of $4\times 10^{-3}$ in the approximation ratio.
There are some exceptions in the case of the TQA strategy optimized with the Nelder-Mead optimizer, where we observe slight improvements in the approximation ratio at shallow depths,
despite the cost reduction (c.f. Fig.~\ref{fig:tqa-results}(c)).

Although we proposed and tested ITLW on QAOA, this idea has a broader view in the sense that it can also be applied in other VQAs,
such as the variational quantum eigensolver (VQE) and the variational circuits in quantum machine learning (QML).
Some details can also be tweaked, for example, the optimization can be parameter-wise (optimize parameter-by-parameter) instead of layer-wise to yield a further reduction in the optimization cost.

\appendix

\section{Details of the initialization strategies}~\label{sec:detail-init}
In this section, we briefly discuss the details of the initialization strategies used in conjunction with ITLW.
We used two different initialization strategies: the bilinear strategy~\cite{bilinear} for depth-progressive (depth-by-depth) initialization,
and the Trotterized quantum annealing (TQA) strategy for direct initialization.

The bilinear strategy takes the optimal parameters from two previous depths $p-1$ and $p-2$ to predict the initial parameters for the current depth $p$.
It leverages the linear-like pattern of the QAOA optimal parameters discussed in many works~\cite{leo2020,Cook2020TheQA}, which resembles the linear annealing schedule. 
Moreover, it is also discovered that the values of the optimal parameters do not stay the same for different depths, coined as the \emph{non-optimality} in~\cite{bilinear}.
Using these properties, the parameters for $p$ can be extended by taking the linear difference between the parameters in $p-1$ and $p-2$.
The idea is to extrapolate the amount of difference between $\bm{\Phi}_{p-1}^*$ and $\bm{\Phi}_{p-2}^*$ to $\bm{\Phi}_{p}$,
where $\bm{\Phi}_{p}$ are the parameters at depth $p$.
For parameters with indices $i\leq p-2$, the parameters are extrapolated using non-optimality:
\begin{equation}
    \begin{split}
        \phi_i^p & = \phi_i^{p-1} + \Delta_i^{p-1,p-2} \\
        & = 2\phi_i^{p-1} - \phi_i^{p-2},
    \end{split}
    \label{eqn:jp-2}
\end{equation}
where the notation $\Delta_{i,j}\equiv \phi_i - \phi_j$. This also applies to the symbols on the superscript.
For the parameters with $i=p-1$, we cannot use Eq.~(\ref{eqn:jp-2}) to extend as the parameter $\phi_{p-1}^{p-2}$ does not exist ($j\leq p$ for $\phi_j^p$),
so we take the difference from the previous index $i=p-2$ instead:
\begin{equation}
    \phi_{p-1}^p = \phi_{p-1}^{p-1} + \Delta_{p-2}^{p-1,p-2}.
    \label{eqn:jp-1}
\end{equation}
For the newly added parameter, we use the linear-like adiabatic pattern to extend:
\begin{equation}
    \begin{split}
        \phi_p^p & = \phi_{p-1}^p + \Delta_{p-1,p-2}^p \\
        & = 2\phi_{p-1}^p - \phi_{p-2}^p.
    \end{split}
    \label{eqn:jp}
\end{equation}
Eq.(\ref{eqn:jp-2}) - (\ref{eqn:jp}) are essentially all the rules required to generate the new initial parameters $\bm{\Phi}_p$. 
The overall procedure for the bilinear strategy is summarized in Algorithm~\ref{alg:bl}.
$\phi_i^p$ denotes the parameter (regardless of $\gamma$ or $\beta$) at depth $p$ with index $i$.

\begin{algorithm}[H]
    \caption{Bilinear initialization}
    \textbf{Input:} Circuit depth $p$, $\bm{\Phi}_{p-1}^*$ and $\bm{\Phi}_{p-2}^*$.
    \begin{algorithmic}[1]
        \For{$i=1...p$}
            \If{$i\leq p-2$}
                \State $\phi_i^p := \phi_i^{p-1} + \Delta_i^{p-1,p-2}$
            \ElsIf{$i=p-1$}
                \State $\phi_i^p := \phi_i^{p-1} + \Delta_{i-1}^{p-1,p-2}$
            \ElsIf{$i=p$}
                \State $\phi_i^p := \phi_{i-1}^p + \Delta_{i-1,i-2}^p$
            \EndIf
        \EndFor
    \end{algorithmic}
    \textbf{Output:} Initial parameters $\bm{\Phi}_p$ for QAOA at depth $p$.
    \label{alg:bl}
\end{algorithm}

The TQA strategy aims to find the ``optimal annealing time'' $T$ for the QAOA to generate the initial parameters, using the angle-index relation: 
$\gamma_i = i\Delta t/p$; $\beta_i=(1-i/p)\Delta t$, where $i$ is the angle index. 
$\Delta t$ is the \emph{annealing time step} and can be related with the \emph{total annealing time} $T$ with the relation $\Delta t = T/p$.
Hence, we get the angle-index relation in terms of $T$:
\begin{equation}
    \gamma_i (T) = \frac{iT}{p^2};\quad \beta_i (T) = \left(1-\frac{i}{p}\right)\frac{T}{p}.
    \label{eqn:tqa}
\end{equation}
Substituting Eq.~(\ref{eqn:tqa}) into (\ref{eqn:fp}) gives the expectation function in terms of $T$. 
Then, the optimal annealing time $T^*$ is found by maximizing $F(T)$:
\begin{equation}
    T^* = \argmax_T F(T).
\end{equation}
$T^*$ is then substitute back into (\ref{eqn:tqa}) to generate the initial angles $(\vgamma(T^*), \vbeta(T^*))$.
The initial angles generated takes a linear form against the angle index $i$.
We summarized the procedure for TQA in Algorithm~\ref{alg:tqa}.

\begin{algorithm}[H]
    \caption{TQA initialization}
    \textbf{Input:} Circuit depth $p$.
    \begin{algorithmic}[1]
        \State Establish the relation between the parameters and the annealing time $T$, and hence $F(T)$:
        \Statex $\gamma_i(T) = iT/p$;
        \Statex $\beta_i(T) = (1-i/p)T/p$;
        \Statex $\expval{H_z} = F(T) = F(\vgamma(T), \vbeta(T))$.
        \State Optimize $F(T)$ to obtain the optimal annealing time $T^*$.
    \end{algorithmic}
    \textbf{Output:} Initial parameters $(\vgamma(T^*), \vbeta(T^*))$.
    \label{alg:tqa}
\end{algorithm}

\bibliography{qaoaref}

\begin{thebibliography}{27}%
\makeatletter
\providecommand \@ifxundefined [1]{%
 \@ifx{#1\undefined}
}%
\providecommand \@ifnum [1]{%
 \ifnum #1\expandafter \@firstoftwo
 \else \expandafter \@secondoftwo
 \fi
}%
\providecommand \@ifx [1]{%
 \ifx #1\expandafter \@firstoftwo
 \else \expandafter \@secondoftwo
 \fi
}%
\providecommand \natexlab [1]{#1}%
\providecommand \enquote  [1]{``#1''}%
\providecommand \bibnamefont  [1]{#1}%
\providecommand \bibfnamefont [1]{#1}%
\providecommand \citenamefont [1]{#1}%
\providecommand \href@noop [0]{\@secondoftwo}%
\providecommand \href [0]{\begingroup \@sanitize@url \@href}%
\providecommand \@href[1]{\@@startlink{#1}\@@href}%
\providecommand \@@href[1]{\endgroup#1\@@endlink}%
\providecommand \@sanitize@url [0]{\catcode `\\12\catcode `\$12\catcode
  `\&12\catcode `\#12\catcode `\^12\catcode `\_12\catcode `\%12\relax}%
\providecommand \@@startlink[1]{}%
\providecommand \@@endlink[0]{}%
\providecommand \url  [0]{\begingroup\@sanitize@url \@url }%
\providecommand \@url [1]{\endgroup\@href {#1}{\urlprefix }}%
\providecommand \urlprefix  [0]{URL }%
\providecommand \Eprint [0]{\href }%
\providecommand \doibase [0]{https://doi.org/}%
\providecommand \selectlanguage [0]{\@gobble}%
\providecommand \bibinfo  [0]{\@secondoftwo}%
\providecommand \bibfield  [0]{\@secondoftwo}%
\providecommand \translation [1]{[#1]}%
\providecommand \BibitemOpen [0]{}%
\providecommand \bibitemStop [0]{}%
\providecommand \bibitemNoStop [0]{.\EOS\space}%
\providecommand \EOS [0]{\spacefactor3000\relax}%
\providecommand \BibitemShut  [1]{\csname bibitem#1\endcsname}%
\let\auto@bib@innerbib\@empty
\bibitem [{\citenamefont {Farhi}\ \emph {et~al.}(2014)\citenamefont {Farhi},
  \citenamefont {Goldstone},\ and\ \citenamefont {Gutmann}}]{farhi2014quantum}%
  \BibitemOpen
  \bibfield  {author} {\bibinfo {author} {\bibfnamefont {E.}~\bibnamefont
  {Farhi}}, \bibinfo {author} {\bibfnamefont {J.}~\bibnamefont {Goldstone}},\
  and\ \bibinfo {author} {\bibfnamefont {S.}~\bibnamefont {Gutmann}},\
  }\href@noop {} {\bibinfo {title} {A quantum approximate optimization
  algorithm}} (\bibinfo {year} {2014}),\ \Eprint
  {https://arxiv.org/abs/arXiv:1411.4028} {arXiv:arXiv:1411.4028 [quant-ph]}
  \BibitemShut {NoStop}%
\bibitem [{\citenamefont {Farhi}\ \emph {et~al.}(2000)\citenamefont {Farhi},
  \citenamefont {Goldstone}, \citenamefont {Gutmann},\ and\ \citenamefont
  {Sipser}}]{farhi:qaa}%
  \BibitemOpen
  \bibfield  {author} {\bibinfo {author} {\bibfnamefont {E.}~\bibnamefont
  {Farhi}}, \bibinfo {author} {\bibfnamefont {J.}~\bibnamefont {Goldstone}},
  \bibinfo {author} {\bibfnamefont {S.}~\bibnamefont {Gutmann}},\ and\ \bibinfo
  {author} {\bibfnamefont {M.}~\bibnamefont {Sipser}},\ }\bibfield  {title}
  {\bibinfo {title} {Quantum computation by adiabatic evolution},\ }\href@noop
  {} {\  (\bibinfo {year} {2000})},\ \Eprint
  {https://arxiv.org/abs/arXiv:quant-ph/0001106} {arXiv:arXiv:quant-ph/0001106}
  \BibitemShut {NoStop}%
\bibitem [{\citenamefont {Lloyd}(2018)}]{lloyd2018quantum}%
  \BibitemOpen
  \bibfield  {author} {\bibinfo {author} {\bibfnamefont {S.}~\bibnamefont
  {Lloyd}},\ }\href@noop {} {\bibinfo {title} {Quantum approximate optimization
  is computationally universal}} (\bibinfo {year} {2018}),\ \Eprint
  {https://arxiv.org/abs/1812.11075} {arXiv:1812.11075 [quant-ph]} \BibitemShut
  {NoStop}%
\bibitem [{\citenamefont {Crooks}(2018)}]{Crooks2018PerformanceOT}%
  \BibitemOpen
  \bibfield  {author} {\bibinfo {author} {\bibfnamefont {G.}~\bibnamefont
  {Crooks}},\ }\bibfield  {title} {\bibinfo {title} {Performance of the quantum
  approximate optimization algorithm on the maximum cut problem},\ }\href@noop
  {} {\bibfield  {journal} {\bibinfo  {journal} {arXiv: Quantum Physics}\ }
  (\bibinfo {year} {2018})}\BibitemShut {NoStop}%
\bibitem [{\citenamefont {Zhou}\ \emph {et~al.}(2020)\citenamefont {Zhou},
  \citenamefont {Wang}, \citenamefont {Choi}, \citenamefont {Pichler},\ and\
  \citenamefont {Lukin}}]{leo2020}%
  \BibitemOpen
  \bibfield  {author} {\bibinfo {author} {\bibfnamefont {L.}~\bibnamefont
  {Zhou}}, \bibinfo {author} {\bibfnamefont {S.-T.}\ \bibnamefont {Wang}},
  \bibinfo {author} {\bibfnamefont {S.}~\bibnamefont {Choi}}, \bibinfo {author}
  {\bibfnamefont {H.}~\bibnamefont {Pichler}},\ and\ \bibinfo {author}
  {\bibfnamefont {M.~D.}\ \bibnamefont {Lukin}},\ }\bibfield  {title} {\bibinfo
  {title} {Quantum approximate optimization algorithm: Performance, mechanism,
  and implementation on near-term devices},\ }\href
  {https://doi.org/10.1103/PhysRevX.10.021067} {\bibfield  {journal} {\bibinfo
  {journal} {Phys. Rev. X}\ }\textbf {\bibinfo {volume} {10}},\ \bibinfo
  {pages} {021067} (\bibinfo {year} {2020})}\BibitemShut {NoStop}%
\bibitem [{\citenamefont {Moussa}\ \emph {et~al.}(2020)\citenamefont {Moussa},
  \citenamefont {Calandra},\ and\ \citenamefont {Dunjko}}]{Moussa_2020}%
  \BibitemOpen
  \bibfield  {author} {\bibinfo {author} {\bibfnamefont {C.}~\bibnamefont
  {Moussa}}, \bibinfo {author} {\bibfnamefont {H.}~\bibnamefont {Calandra}},\
  and\ \bibinfo {author} {\bibfnamefont {V.}~\bibnamefont {Dunjko}},\
  }\bibfield  {title} {\bibinfo {title} {To quantum or not to quantum: towards
  algorithm selection in near-term quantum optimization},\ }\href
  {https://doi.org/10.1088/2058-9565/abb8e5} {\bibfield  {journal} {\bibinfo
  {journal} {Quantum Science and Technology}\ }\textbf {\bibinfo {volume}
  {5}},\ \bibinfo {pages} {044009} (\bibinfo {year} {2020})}\BibitemShut
  {NoStop}%
\bibitem [{\citenamefont {Marwaha}(2021)}]{Marwaha_2021}%
  \BibitemOpen
  \bibfield  {author} {\bibinfo {author} {\bibfnamefont {K.}~\bibnamefont
  {Marwaha}},\ }\bibfield  {title} {\bibinfo {title} {Local classical
  {MAX}-{CUT} algorithm outperforms $p=2$ {QAOA} on high-girth regular
  graphs},\ }\href {https://doi.org/10.22331/q-2021-04-20-437} {\bibfield
  {journal} {\bibinfo  {journal} {Quantum}\ }\textbf {\bibinfo {volume} {5}},\
  \bibinfo {pages} {437} (\bibinfo {year} {2021})}\BibitemShut {NoStop}%
\bibitem [{\citenamefont {Hadfield}\ \emph {et~al.}(2019)\citenamefont
  {Hadfield}, \citenamefont {Wang}, \citenamefont {O{\textquotesingle}Gorman},
  \citenamefont {Rieffel}, \citenamefont {Venturelli},\ and\ \citenamefont
  {Biswas}}]{new-qaoa}%
  \BibitemOpen
  \bibfield  {author} {\bibinfo {author} {\bibfnamefont {S.}~\bibnamefont
  {Hadfield}}, \bibinfo {author} {\bibfnamefont {Z.}~\bibnamefont {Wang}},
  \bibinfo {author} {\bibfnamefont {B.}~\bibnamefont
  {O{\textquotesingle}Gorman}}, \bibinfo {author} {\bibfnamefont
  {E.}~\bibnamefont {Rieffel}}, \bibinfo {author} {\bibfnamefont
  {D.}~\bibnamefont {Venturelli}},\ and\ \bibinfo {author} {\bibfnamefont
  {R.}~\bibnamefont {Biswas}},\ }\bibfield  {title} {\bibinfo {title} {From the
  quantum approximate optimization algorithm to a quantum alternating operator
  ansatz},\ }\href {https://doi.org/10.3390/a12020034} {\bibfield  {journal}
  {\bibinfo  {journal} {Algorithms}\ }\textbf {\bibinfo {volume} {12}},\
  \bibinfo {pages} {34} (\bibinfo {year} {2019})}\BibitemShut {NoStop}%
\bibitem [{\citenamefont {Zhu}\ \emph {et~al.}(2022)\citenamefont {Zhu},
  \citenamefont {Tang}, \citenamefont {Barron}, \citenamefont
  {Calderon-Vargas}, \citenamefont {Mayhall}, \citenamefont {Barnes},\ and\
  \citenamefont {Economou}}]{adapt-qaoa}%
  \BibitemOpen
  \bibfield  {author} {\bibinfo {author} {\bibfnamefont {L.}~\bibnamefont
  {Zhu}}, \bibinfo {author} {\bibfnamefont {H.~L.}\ \bibnamefont {Tang}},
  \bibinfo {author} {\bibfnamefont {G.~S.}\ \bibnamefont {Barron}}, \bibinfo
  {author} {\bibfnamefont {F.~A.}\ \bibnamefont {Calderon-Vargas}}, \bibinfo
  {author} {\bibfnamefont {N.~J.}\ \bibnamefont {Mayhall}}, \bibinfo {author}
  {\bibfnamefont {E.}~\bibnamefont {Barnes}},\ and\ \bibinfo {author}
  {\bibfnamefont {S.~E.}\ \bibnamefont {Economou}},\ }\bibfield  {title}
  {\bibinfo {title} {Adaptive quantum approximate optimization algorithm for
  solving combinatorial problems on a quantum computer},\ }\href
  {https://doi.org/10.1103/PhysRevResearch.4.033029} {\bibfield  {journal}
  {\bibinfo  {journal} {Phys. Rev. Res.}\ }\textbf {\bibinfo {volume} {4}},\
  \bibinfo {pages} {033029} (\bibinfo {year} {2022})}\BibitemShut {NoStop}%
\bibitem [{\citenamefont {Bartschi}\ and\ \citenamefont
  {Eidenbenz}(2020)}]{grover-mixer}%
  \BibitemOpen
  \bibfield  {author} {\bibinfo {author} {\bibfnamefont {A.}~\bibnamefont
  {Bartschi}}\ and\ \bibinfo {author} {\bibfnamefont {S.}~\bibnamefont
  {Eidenbenz}},\ }\bibfield  {title} {\bibinfo {title} {Grover mixers for
  {QAOA}: Shifting complexity from mixer design to state preparation},\ }in\
  \href {https://doi.org/10.1109/qce49297.2020.00020} {\emph {\bibinfo
  {booktitle} {2020 {IEEE} International Conference on Quantum Computing and
  Engineering ({QCE})}}}\ (\bibinfo  {publisher} {{IEEE}},\ \bibinfo {year}
  {2020})\BibitemShut {NoStop}%
\bibitem [{\citenamefont {Herrman}\ \emph {et~al.}(2021)\citenamefont
  {Herrman}, \citenamefont {Lotshaw}, \citenamefont {Ostrowski}, \citenamefont
  {Humble},\ and\ \citenamefont {Siopsis}}]{ma-qaoa}%
  \BibitemOpen
  \bibfield  {author} {\bibinfo {author} {\bibfnamefont {R.}~\bibnamefont
  {Herrman}}, \bibinfo {author} {\bibfnamefont {P.~C.}\ \bibnamefont
  {Lotshaw}}, \bibinfo {author} {\bibfnamefont {J.}~\bibnamefont {Ostrowski}},
  \bibinfo {author} {\bibfnamefont {T.~S.}\ \bibnamefont {Humble}},\ and\
  \bibinfo {author} {\bibfnamefont {G.}~\bibnamefont {Siopsis}},\ }\href@noop
  {} {\bibinfo {title} {Multi-angle quantum approximate optimization
  algorithm}} (\bibinfo {year} {2021}),\ \Eprint
  {https://arxiv.org/abs/2109.11455} {arXiv:2109.11455 [quant-ph]} \BibitemShut
  {NoStop}%
\bibitem [{\citenamefont {Grant}\ \emph {et~al.}(2019)\citenamefont {Grant},
  \citenamefont {Wossnig}, \citenamefont {Ostaszewski},\ and\ \citenamefont
  {Benedetti}}]{Grant_2019}%
  \BibitemOpen
  \bibfield  {author} {\bibinfo {author} {\bibfnamefont {E.}~\bibnamefont
  {Grant}}, \bibinfo {author} {\bibfnamefont {L.}~\bibnamefont {Wossnig}},
  \bibinfo {author} {\bibfnamefont {M.}~\bibnamefont {Ostaszewski}},\ and\
  \bibinfo {author} {\bibfnamefont {M.}~\bibnamefont {Benedetti}},\ }\bibfield
  {title} {\bibinfo {title} {An initialization strategy for addressing barren
  plateaus in parametrized quantum circuits},\ }\href
  {https://doi.org/10.22331/q-2019-12-09-214} {\bibfield  {journal} {\bibinfo
  {journal} {Quantum}\ }\textbf {\bibinfo {volume} {3}},\ \bibinfo {pages}
  {214} (\bibinfo {year} {2019})}\BibitemShut {NoStop}%
\bibitem [{\citenamefont {Uvarov}\ and\ \citenamefont
  {Biamonte}(2021)}]{barren_vqa2021}%
  \BibitemOpen
  \bibfield  {author} {\bibinfo {author} {\bibfnamefont {A.~V.}\ \bibnamefont
  {Uvarov}}\ and\ \bibinfo {author} {\bibfnamefont {J.~D.}\ \bibnamefont
  {Biamonte}},\ }\bibfield  {title} {\bibinfo {title} {On barren plateaus and
  cost function locality in variational quantum algorithms},\ }\href
  {https://doi.org/10.1088/1751-8121/abfac7} {\bibfield  {journal} {\bibinfo
  {journal} {Journal of Physics A: Mathematical and Theoretical}\ }\textbf
  {\bibinfo {volume} {54}},\ \bibinfo {pages} {245301} (\bibinfo {year}
  {2021})}\BibitemShut {NoStop}%
\bibitem [{\citenamefont {Cerezo}\ \emph {et~al.}(2021)\citenamefont {Cerezo},
  \citenamefont {Sone}, \citenamefont {Volkoff}, \citenamefont {Cincio},\ and\
  \citenamefont {Coles}}]{Cerezo_2021}%
  \BibitemOpen
  \bibfield  {author} {\bibinfo {author} {\bibfnamefont {M.}~\bibnamefont
  {Cerezo}}, \bibinfo {author} {\bibfnamefont {A.}~\bibnamefont {Sone}},
  \bibinfo {author} {\bibfnamefont {T.}~\bibnamefont {Volkoff}}, \bibinfo
  {author} {\bibfnamefont {L.}~\bibnamefont {Cincio}},\ and\ \bibinfo {author}
  {\bibfnamefont {P.~J.}\ \bibnamefont {Coles}},\ }\bibfield  {title} {\bibinfo
  {title} {Cost function dependent barren plateaus in shallow parametrized
  quantum circuits},\ }\bibfield  {journal} {\bibinfo  {journal} {Nature
  Communications}\ }\textbf {\bibinfo {volume} {12}},\ \href
  {https://doi.org/10.1038/s41467-021-21728-w} {10.1038/s41467-021-21728-w}
  (\bibinfo {year} {2021})\BibitemShut {NoStop}%
\bibitem [{\citenamefont {Wang}\ \emph {et~al.}(2021)\citenamefont {Wang},
  \citenamefont {Fontana}, \citenamefont {Cerezo}, \citenamefont {Sharma},
  \citenamefont {Sone}, \citenamefont {Cincio},\ and\ \citenamefont
  {Coles}}]{Wang_2021}%
  \BibitemOpen
  \bibfield  {author} {\bibinfo {author} {\bibfnamefont {S.}~\bibnamefont
  {Wang}}, \bibinfo {author} {\bibfnamefont {E.}~\bibnamefont {Fontana}},
  \bibinfo {author} {\bibfnamefont {M.}~\bibnamefont {Cerezo}}, \bibinfo
  {author} {\bibfnamefont {K.}~\bibnamefont {Sharma}}, \bibinfo {author}
  {\bibfnamefont {A.}~\bibnamefont {Sone}}, \bibinfo {author} {\bibfnamefont
  {L.}~\bibnamefont {Cincio}},\ and\ \bibinfo {author} {\bibfnamefont {P.~J.}\
  \bibnamefont {Coles}},\ }\bibfield  {title} {\bibinfo {title} {Noise-induced
  barren plateaus in variational quantum algorithms},\ }\bibfield  {journal}
  {\bibinfo  {journal} {Nature Communications}\ }\textbf {\bibinfo {volume}
  {12}},\ \href {https://doi.org/10.1038/s41467-021-27045-6}
  {10.1038/s41467-021-27045-6} (\bibinfo {year} {2021})\BibitemShut {NoStop}%
\bibitem [{\citenamefont {Guerreschi}\ and\ \citenamefont
  {Matsuura}(2019)}]{Guerreschi_2019}%
  \BibitemOpen
  \bibfield  {author} {\bibinfo {author} {\bibfnamefont {G.~G.}\ \bibnamefont
  {Guerreschi}}\ and\ \bibinfo {author} {\bibfnamefont {A.~Y.}\ \bibnamefont
  {Matsuura}},\ }\bibfield  {title} {\bibinfo {title} {Qaoa for max-cut
  requires hundreds of qubits for quantum speed-up},\ }\bibfield  {journal}
  {\bibinfo  {journal} {Scientific Reports}\ }\textbf {\bibinfo {volume} {9}},\
  \href {https://doi.org/10.1038/s41598-019-43176-9}
  {10.1038/s41598-019-43176-9} (\bibinfo {year} {2019})\BibitemShut {NoStop}%
\bibitem [{\citenamefont {Sack}\ \emph {et~al.}(2023)\citenamefont {Sack},
  \citenamefont {Medina}, \citenamefont {Kueng},\ and\ \citenamefont
  {Serbyn}}]{sack-greedy2023}%
  \BibitemOpen
  \bibfield  {author} {\bibinfo {author} {\bibfnamefont {S.~H.}\ \bibnamefont
  {Sack}}, \bibinfo {author} {\bibfnamefont {R.~A.}\ \bibnamefont {Medina}},
  \bibinfo {author} {\bibfnamefont {R.}~\bibnamefont {Kueng}},\ and\ \bibinfo
  {author} {\bibfnamefont {M.}~\bibnamefont {Serbyn}},\ }\bibfield  {title}
  {\bibinfo {title} {Recursive greedy initialization of the quantum approximate
  optimization algorithm with guaranteed improvement},\ }\href
  {https://doi.org/10.1103/PhysRevA.107.062404} {\bibfield  {journal} {\bibinfo
   {journal} {Phys. Rev. A}\ }\textbf {\bibinfo {volume} {107}},\ \bibinfo
  {pages} {062404} (\bibinfo {year} {2023})}\BibitemShut {NoStop}%
\bibitem [{\citenamefont {Sack}\ and\ \citenamefont
  {Serbyn}(2021)}]{sack2021quantum}%
  \BibitemOpen
  \bibfield  {author} {\bibinfo {author} {\bibfnamefont {S.~H.}\ \bibnamefont
  {Sack}}\ and\ \bibinfo {author} {\bibfnamefont {M.}~\bibnamefont {Serbyn}},\
  }\href@noop {} {\bibinfo {title} {Quantum annealing initialization of the
  quantum approximate optimization algorithm}} (\bibinfo {year} {2021}),\
  \Eprint {https://arxiv.org/abs/2101.05742} {arXiv:2101.05742 [quant-ph]}
  \BibitemShut {NoStop}%
\bibitem [{\citenamefont {Lee}\ \emph {et~al.}(2021)\citenamefont {Lee},
  \citenamefont {Saito}, \citenamefont {Cai},\ and\ \citenamefont
  {Asai}}]{lee2021parameters}%
  \BibitemOpen
  \bibfield  {author} {\bibinfo {author} {\bibfnamefont {X.}~\bibnamefont
  {Lee}}, \bibinfo {author} {\bibfnamefont {Y.}~\bibnamefont {Saito}}, \bibinfo
  {author} {\bibfnamefont {D.}~\bibnamefont {Cai}},\ and\ \bibinfo {author}
  {\bibfnamefont {N.}~\bibnamefont {Asai}},\ }\bibfield  {title} {\bibinfo
  {title} {Parameters fixing strategy for quantum approximate optimization
  algorithm},\ }in\ \href {https://doi.org/10.1109/qce52317.2021.00016} {\emph
  {\bibinfo {booktitle} {2021 {IEEE} International Conference on Quantum
  Computing and Engineering ({QCE})}}}\ (\bibinfo  {publisher} {{IEEE}},\
  \bibinfo {year} {2021})\BibitemShut {NoStop}%
\bibitem [{\citenamefont {Shaydulin}\ \emph {et~al.}(2019)\citenamefont
  {Shaydulin}, \citenamefont {Safro},\ and\ \citenamefont
  {Larson}}]{multistart}%
  \BibitemOpen
  \bibfield  {author} {\bibinfo {author} {\bibfnamefont {R.}~\bibnamefont
  {Shaydulin}}, \bibinfo {author} {\bibfnamefont {I.}~\bibnamefont {Safro}},\
  and\ \bibinfo {author} {\bibfnamefont {J.}~\bibnamefont {Larson}},\
  }\bibfield  {title} {\bibinfo {title} {Multistart methods for quantum
  approximate optimization},\ }\bibfield  {journal} {\bibinfo  {journal} {2019
  IEEE High Performance Extreme Computing Conference (HPEC)}\ }\href
  {https://doi.org/10.1109/hpec.2019.8916288} {10.1109/hpec.2019.8916288}
  (\bibinfo {year} {2019})\BibitemShut {NoStop}%
\bibitem [{\citenamefont {Campos}\ \emph {et~al.}(2021)\citenamefont {Campos},
  \citenamefont {Rabinovich}, \citenamefont {Akshay},\ and\ \citenamefont
  {Biamonte}}]{campos2021}%
  \BibitemOpen
  \bibfield  {author} {\bibinfo {author} {\bibfnamefont {E.}~\bibnamefont
  {Campos}}, \bibinfo {author} {\bibfnamefont {D.}~\bibnamefont {Rabinovich}},
  \bibinfo {author} {\bibfnamefont {V.}~\bibnamefont {Akshay}},\ and\ \bibinfo
  {author} {\bibfnamefont {J.}~\bibnamefont {Biamonte}},\ }\bibfield  {title}
  {\bibinfo {title} {Training saturation in layerwise quantum approximate
  optimization},\ }\bibfield  {journal} {\bibinfo  {journal} {Physical Review
  A}\ }\textbf {\bibinfo {volume} {104}},\ \href
  {https://doi.org/10.1103/physreva.104.l030401} {10.1103/physreva.104.l030401}
  (\bibinfo {year} {2021})\BibitemShut {NoStop}%
\bibitem [{\citenamefont {Skolik}\ \emph {et~al.}(2021)\citenamefont {Skolik},
  \citenamefont {McClean}, \citenamefont {Mohseni}, \citenamefont {van~der
  Smagt},\ and\ \citenamefont {Leib}}]{lw-qnn}%
  \BibitemOpen
  \bibfield  {author} {\bibinfo {author} {\bibfnamefont {A.}~\bibnamefont
  {Skolik}}, \bibinfo {author} {\bibfnamefont {J.~R.}\ \bibnamefont {McClean}},
  \bibinfo {author} {\bibfnamefont {M.}~\bibnamefont {Mohseni}}, \bibinfo
  {author} {\bibfnamefont {P.}~\bibnamefont {van~der Smagt}},\ and\ \bibinfo
  {author} {\bibfnamefont {M.}~\bibnamefont {Leib}},\ }\bibfield  {title}
  {\bibinfo {title} {Layerwise learning for quantum neural networks},\
  }\bibfield  {journal} {\bibinfo  {journal} {Quantum Machine Intelligence}\
  }\textbf {\bibinfo {volume} {3}},\ \href
  {https://doi.org/10.1007/s42484-020-00036-4} {10.1007/s42484-020-00036-4}
  (\bibinfo {year} {2021})\BibitemShut {NoStop}%
\bibitem [{\citenamefont {Lee}\ \emph {et~al.}(2023)\citenamefont {Lee},
  \citenamefont {Xie}, \citenamefont {Cai}, \citenamefont {Saito},\ and\
  \citenamefont {Asai}}]{bilinear}%
  \BibitemOpen
  \bibfield  {author} {\bibinfo {author} {\bibfnamefont {X.}~\bibnamefont
  {Lee}}, \bibinfo {author} {\bibfnamefont {N.}~\bibnamefont {Xie}}, \bibinfo
  {author} {\bibfnamefont {D.}~\bibnamefont {Cai}}, \bibinfo {author}
  {\bibfnamefont {Y.}~\bibnamefont {Saito}},\ and\ \bibinfo {author}
  {\bibfnamefont {N.}~\bibnamefont {Asai}},\ }\bibfield  {title} {\bibinfo
  {title} {A depth-progressive initialization strategy for quantum approximate
  optimization algorithm},\ }\href {https://doi.org/10.3390/math11092176}
  {\bibfield  {journal} {\bibinfo  {journal} {Mathematics}\ }\textbf {\bibinfo
  {volume} {11}},\ \bibinfo {pages} {2176} (\bibinfo {year}
  {2023})}\BibitemShut {NoStop}%
\bibitem [{\citenamefont {Lotshaw}\ \emph {et~al.}(2021)\citenamefont
  {Lotshaw}, \citenamefont {Humble}, \citenamefont {Herrman}, \citenamefont
  {Ostrowski},\ and\ \citenamefont {Siopsis}}]{Lotshaw_2021}%
  \BibitemOpen
  \bibfield  {author} {\bibinfo {author} {\bibfnamefont {P.~C.}\ \bibnamefont
  {Lotshaw}}, \bibinfo {author} {\bibfnamefont {T.~S.}\ \bibnamefont {Humble}},
  \bibinfo {author} {\bibfnamefont {R.}~\bibnamefont {Herrman}}, \bibinfo
  {author} {\bibfnamefont {J.}~\bibnamefont {Ostrowski}},\ and\ \bibinfo
  {author} {\bibfnamefont {G.}~\bibnamefont {Siopsis}},\ }\bibfield  {title}
  {\bibinfo {title} {Empirical performance bounds for quantum approximate
  optimization},\ }\bibfield  {journal} {\bibinfo  {journal} {Quantum
  Information Processing}\ }\textbf {\bibinfo {volume} {20}},\ \href
  {https://doi.org/10.1007/s11128-021-03342-3} {10.1007/s11128-021-03342-3}
  (\bibinfo {year} {2021})\BibitemShut {NoStop}%
\bibitem [{\citenamefont {Morales}\ and\ \citenamefont
  {Nocedal}(2011)}]{l-bfgs-b}%
  \BibitemOpen
  \bibfield  {author} {\bibinfo {author} {\bibfnamefont {J.~L.}\ \bibnamefont
  {Morales}}\ and\ \bibinfo {author} {\bibfnamefont {J.}~\bibnamefont
  {Nocedal}},\ }\bibfield  {title} {\bibinfo {title} {Remark on “algorithm
  778: L-bfgs-b: Fortran subroutines for large-scale bound constrained
  optimization”},\ }\bibfield  {journal} {\bibinfo  {journal} {ACM Trans.
  Math. Softw.}\ }\textbf {\bibinfo {volume} {38}},\ \href
  {https://doi.org/10.1145/2049662.2049669} {10.1145/2049662.2049669} (\bibinfo
  {year} {2011})\BibitemShut {NoStop}%
\bibitem [{\citenamefont {Nelder}\ and\ \citenamefont
  {Mead}(1965)}]{neldermead}%
  \BibitemOpen
  \bibfield  {author} {\bibinfo {author} {\bibfnamefont {J.~A.}\ \bibnamefont
  {Nelder}}\ and\ \bibinfo {author} {\bibfnamefont {R.}~\bibnamefont {Mead}},\
  }\bibfield  {title} {\bibinfo {title} {{A Simplex Method for Function
  Minimization}},\ }\href {https://doi.org/10.1093/comjnl/7.4.308} {\bibfield
  {journal} {\bibinfo  {journal} {The Computer Journal}\ }\textbf {\bibinfo
  {volume} {7}},\ \bibinfo {pages} {308} (\bibinfo {year} {1965})},\ \Eprint
  {https://arxiv.org/abs/https://academic.oup.com/comjnl/article-pdf/7/4/308/1013182/7-4-308.pdf}
  {https://academic.oup.com/comjnl/article-pdf/7/4/308/1013182/7-4-308.pdf}
  \BibitemShut {NoStop}%
\bibitem [{\citenamefont {Cook}\ \emph {et~al.}(2020)\citenamefont {Cook},
  \citenamefont {Eidenbenz},\ and\ \citenamefont
  {B{\"a}rtschi}}]{Cook2020TheQA}%
  \BibitemOpen
  \bibfield  {author} {\bibinfo {author} {\bibfnamefont {J.}~\bibnamefont
  {Cook}}, \bibinfo {author} {\bibfnamefont {S.}~\bibnamefont {Eidenbenz}},\
  and\ \bibinfo {author} {\bibfnamefont {A.}~\bibnamefont {B{\"a}rtschi}},\
  }\bibfield  {title} {\bibinfo {title} {The quantum alternating operator
  ansatz on maximum k-vertex cover},\ }\href@noop {} {\bibfield  {journal}
  {\bibinfo  {journal} {2020 IEEE International Conference on Quantum Computing
  and Engineering (QCE)}\ ,\ \bibinfo {pages} {83}} (\bibinfo {year}
  {2020})}\BibitemShut {NoStop}%
\end{thebibliography}%
\end{document}